\documentclass[usegraphicx,useAMS,usenatbib,fleqn]{mn2e}
\include{latex_macros}
\usepackage{float}

\usepackage{times}
\usepackage{amsmath}
\usepackage{amsfonts}
\title[Inflated radii in PMS stars]{The Gaia-ESO Survey: Lithium depletion in the Gamma Velorum cluster
  and inflated radii in low-mass pre-main-sequence stars}

\author[R. D. Jeffries et al.]
  {R. D.~Jeffries$^1$\thanks{r.d.jeffries@keele.ac.uk}, R. J.~Jackson$^1$, E. Franciosini$^2$,
    S. Randich$^2$, D. Barrado$^3$, A. Frasca$^4$, \newauthor
    A. Klutsch$^4$, A.~C. Lanzafame$^{4,5}$,
    L. Prisinzano$^6$, G.~G. Sacco$^2$, G. Gilmore$^7$, A. Vallenari$^{8}$,
    \newauthor E.~J. Alfaro$^9$, S.~E. Koposov$^{7,10}$, E. Pancino$^{2,11}$,
    A. Bayo$^{12}$, A.~R. Casey$^7$, M.~T. Costado$^9$, 
    \newauthor F. Damiani$^6$, 
    A. Hourihane$^7$, J. Lewis$^7$,
    P. Jofre$^{7,13}$, L. Magrini$^2$, L. Monaco$^{14}$,
    L. Morbidelli$^2$, \newauthor C.~C. Worley$^7$,
    S. Zaggia$^8$ and
    T. Zwitter$^{15}$\\
  $^1$ Astrophysics Group, Keele University, Keele, 
      Staffordshire ST5 5BG, United Kingdom\\
$^2$ INAF - Osservatorio Astrofisico di Arcetri, Largo E. Fermi 5, 50125, Florence, Italy\\
$^3$ Depto. Astrofisica, Centro de Astrobiologia (INTA-CSIC), ESAC
      campus, Camino Bajo del Castillo s/n, E-28692 Villanueva de la
      Ca\~nada, Spain\\
$^4$ INAF - Osservatorio Astrofisico di Catania, via S. Sofia 78, 95123, Catania, Italy\\
$^5$ Dipartimento di Fisica e Astronomia, Sezione Astrofisica, Universit\'a di Catania, via S. Sofia 78, 95123, Catania, Italy\\
$^6$ INAF - Osservatorio Astronomico di Palermo, Piazza del Parlamento 1, 90134, Palermo, Italy\\
$^7$ Institute of Astronomy, University of Cambridge, Madingley Road,
      Cambridge CB3 0HA, United Kingdom\\
$^8$ INAF - Padova Observatory, Vicolo dell'Osservatorio 5, 35122
      Padova, Italy\\
$^9$ Instituto de Astrof\'isica de Andaluc\'{i}a-CSIC, Apdo. 3004,
      18080, Granada, Spain\\
$^{10}$ Moscow MV Lomonosov State University, Sternberg Astronomical Institute, Moscow 119992, Russia\\
$^{11}$ ASI Science Data Center, Via del Politecnico SNC, 00133 Roma, Italy\\
$^{12}$ Instituto de F\'isica y Astronomi\'ia, Universidad de Valparai\'iso, Chile\\
$^{13}$ N\'ucleo de Astronom\'ia, Facultad de Ingenier\'ia, Universidad Diego Portales,  Av. Ejercito 441, Santiago, Chile\\
$^{14}$ Departamento de Ciencias Fisicas, Universidad Andres Bello,
      Republica 220, Santiago, Chile\\
$^{15}$ Faculty of Mathematics and Physics, University of Ljubljana, Jadranska 19, 1000, Ljubljana, Slovenia
}
\date{19/09/16}

\pagerange{\pageref{firstpage}--\pageref{lastpage}} \pubyear{2016}

\def\LaTeX{L\kern-.36em\raise.3ex\hbox{a}\kern-.15em
    T\kern-.1667em\lower.7ex\hbox{E}\kern-.125emX}

\begin{document}
\label{firstpage}
\maketitle

\begin{abstract}
We show that
non-magnetic models for the evolution of pre-main-sequence (PMS)
 stars {\it cannot} simultaneously describe the
colour-magnitude diagram (CMD) and the pattern of lithium depletion seen in
the cluster of young, low-mass stars surrounding 
$\gamma^2$ Velorum. The age of $7.5 \pm 1$\,Myr inferred from the CMD is much
younger than that implied by the strong Li depletion seen in the
cluster M-dwarfs and the Li depletion
occurs at much redder colours than predicted. The epoch at
which a star of a given mass depletes its Li and the surface
temperature of that star are both dependent on its radius. We demonstrate that
if the low-mass stars have radii $\sim 10$ per cent larger at a given
mass and age, then both the CMD and Li depletion pattern of the Gamma
Vel cluster are explained at a common age of $\simeq 18$--21\,Myr.
This radius inflation could be produced by some combination of magnetic suppression
of convection and extensive cool starspots. Models that incorporate
radius inflation suggest that PMS stars similar to those in the Gamma
Vel cluster,
in the range $0.2<M/M_{\odot}<0.7$, 
are at least a factor of two older and 
$\sim 7$ per cent cooler than previously thought and that their masses
are much larger (by $>30$ per cent) than inferred from conventional,
non-magnetic models in the Hertzsprung-Russell diagram.
Systematic changes of this size may be of great importance in
understanding the evolution of young stars, disc lifetimes and the
formation of planetary systems.
\end{abstract}


\section{Introduction}
Precise measurements of main-sequence K- and M-dwarf radii in eclipsing
binaries have revealed alarming discrepancies between theoretical
models and observations. For a given mass, the absolute radii of stars
with $0.2<M/M_{\odot}<0.8$ can be $\sim 10$ per cent larger than
predicted and hence, for a given luminosity, the effective temperature
is overestimated by $\sim 5$ per cent (e.g. Lopez-Morales \& Ribas
2005; Morales et al. 2009a,b; Torres 2013). Since interferometric radii
for nearby, slowly-rotating low-mass stars are much closer
to ``standard'', non-magnetic evolutionary models (e.g. Demory et
al. 2009; Boyajian et al. 2012), it is possible that the radius
inflation of fast-rotating binary components is explained by
dynamo-generated magnetic activity (Morales, Ribas \& Jordi 2008). The
exact mechanism is debated, but could be magnetic fields
inhibiting convection throughout the star (Mullan \& MacDonald 2001;
Feiden \& Chaboyer 2013, 2014) or cool, magnetic starspots that block
outward flux at the stellar surface (MacDonald \& Mullan 2013; Jackson
\& Jeffries 2014b).

\nocite{lopez-morales05a}
\nocite{morales09a}
\nocite{morales09b}
\nocite{torres13a}
\nocite{demory09a}
\nocite{boyajian12a}
\nocite{morales08a}
\nocite{mullan01a}
\nocite{feiden13a}
\nocite{feiden14a}
\nocite{macdonald13a}
\nocite{jackson14a}

If magnetic activity is implicated in inflating radii, then the same effect
should also be present in low-mass pre main sequence (PMS) and zero-age main
sequence (ZAMS) stars, which are frequently found to be as fast-rotating and
magnetically active as the tidally-locked components of older eclipsing
binary systems. Besides revealing an interesting new facet of the
astrophysics of low-mass stars, radius inflation in PMS stars would have important
practical consequences. Models that include such inflation lead to
substantial increases in both the ages and masses that would be 
inferred from colour-magnitude diagrams (CMDs) and
Hertzsprung-Russell diagrams (HRDs; perhaps by factors of two or more -- see
Somers \& Pinsonneault 2015b; Feiden 2016), systematically affecting
studies of the star formation process and the early evolution of stars
and their planetary systems by altering the inferred timescales of
important phases (Bell et al. 2013; Soderblom et al. 2014).

At present it is not possible to directly measure the masses, radii
{\it and} ages of young low-mass stars in order to compare them with
evolutionary models. However, it has been suggested that discrepancies
between masses, radii, temperatures and luminosities in PMS binaries
(Kraus et al. 2015. 2016),
the lithium depletion dispersion in PMS stars (Somers \& Pinsonneault
2015a,b)  and the anomalous colours
of PMS and ZAMS stars with respect to model isochrones (Stauffer
et al. 2003; Covey et al. 2016 and see Section 5), might be explained by
radius inflation or magnetic activity.  Indirect evidence for enlarged
stellar radii in young clusters, based on the product of their rotation
periods and projected equatorial velocities, supports this view
(Jackson, Jeffries \& Maxted 2009; Jackson et al. 2016).

\nocite{kraus15a}
\nocite{david16a}
\nocite{hillenbrand04a}
\nocite{azulay15a}
\nocite{rizzuto16a}
\nocite{stauffer03a}
\nocite{covey16a}
\nocite{jackson16a}
\nocite{jackson09a}
\nocite{jackson14b}
\nocite{somers14a}
\nocite{somers15a}

\nocite{somers15b}
\nocite{feiden16a}
\nocite{bell13a}
\nocite{soderblom14a}

The Gaia-ESO spectroscopic survey of representative stellar populations
at the ESO Very Large Telescope (GES, Gilmore et al. 2012; Randich \&
Gilmore 2013) includes observations of large, unbiased samples of PMS
stars in young clusters, hence providing an important dataset to
further investigate these issues.  This paper presents the results of a
test, first suggested by Yee \& Jensen (2010, see also Feiden 2016 and
Messina et al. 2016), that exploits the sensitive mass and radius
dependence of the onset of lithium depletion in fully convective PMS
stars.  Using GES observations of lithium, we find strong evidence for
radius inflation in a large group of low-mass PMS stars belonging to
the Gamma Vel cluster (Jeffries et al. 2009). We show that standard,
non-magnetic models fail to simultaneously describe the CMD and the
pattern of lithium depletion, but that a simple inflation of the radius
at a given mass and age could solve this problem.

\nocite{jeffries09a}
\nocite{jeffries14a}
\nocite{prisinzano16a}
\nocite{yee10a}
\nocite{messina16a}

 \section{The Gamma Vel cluster and non-magnetic models} \label{gammavel}

\begin{figure*}
\centering
\includegraphics[width=160mm]{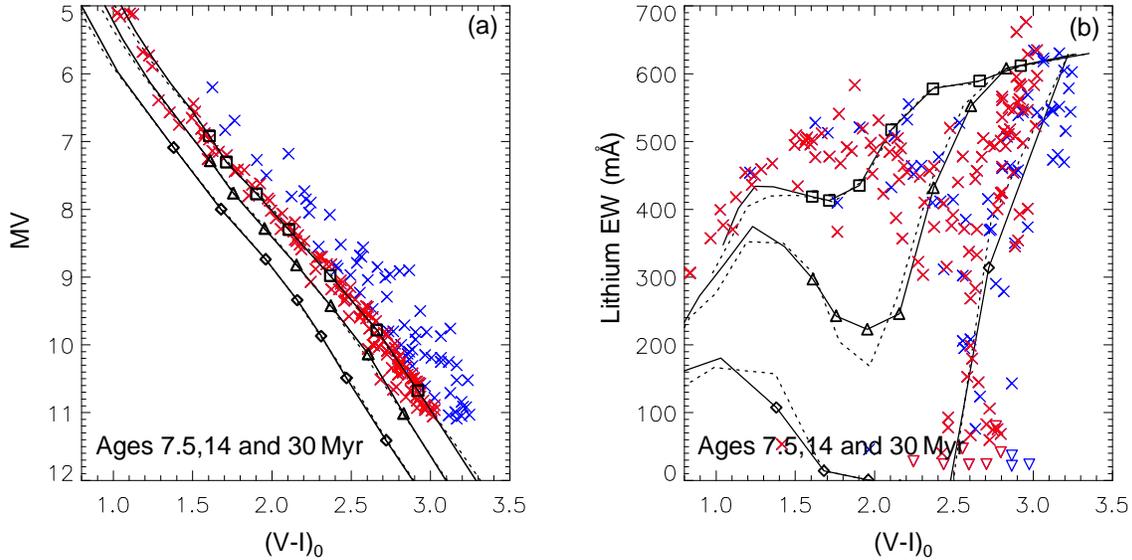}
\caption{The absolute 
  magnitude versus intrinsic colour diagram and lithium depletion pattern
  in the Gamma Vel cluster. The left hand panel shows the absolute $V$
  magnitude versus intrinsic $(V-I)_0$
  CMD for a set of
  members defined in Prisinzano et al. (2016). Overplotted are isochrones
  from the Baraffe et al. (2015, solid) and Dotter et al. (2008,
  dashed) isochrones at 7.5 (diamonds), 14 (triangles) and 30 Myr
  (squares). A distance
  modulus of 7.76 mag, $A_V=0.13$ mag and $E(V-I)=0.055$ mag have been
  assumed.  Mass points
  from $0.2\,M_{\odot}$ to $0.8\,M_{\odot}$ in $0.1\,M_{\odot}$ steps
  are indicated by the open symbols on each isochrone. The right hand panel shows
  Li~6708\AA\ equivalent width versus colour for the same cluster
  members, with predicted isochrones from the same set of models and at
  the same ages as the left hand panel. The blue points in each plot
  mark possible unresolved binaries, chosen as the third of points
  furthest away from the best-fitting isochrone in the CMD. The median
  uncertainty in EWLi is 15\,m\AA. Downward pointing triangles indicate
upper limits.}
\end{figure*}

The Gamma Vel cluster is a group of several hundred, low-mass PMS stars
and some tens of higher mass ZAMS stars, that are
spatially concentrated around the massive Wolf-Rayet binary system
$\gamma^2$ Vel (Pozzo et al. 2000; Jeffries et al. 2009). 
The cluster was one of the first targets for the GES; 
intermediate-resolution ($R\simeq 17\,000$) 
fibre spectroscopy in the cluster, taken with the Giraffe spectrograph and HR15N
grating ($\lambda\lambda$ 6444-6816\AA), has been extensively
described and analysed in Jeffries et al. (2014) and Prisinzano et
al. (2016), establishing a kinematically coherent cluster of low-mass PMS stars
at an age of $\sim 10$ Myr. The cluster appears to have two kinematic
components, separated by just 2 km\,s$^{-1}$, that are almost
indistinguishable in terms of their positions in CMDs
and the lithium depletion seen among their low-mass members.
No distinction is made between these subgroups in this paper.

\nocite{pozzo00a}
\nocite{gilmore12a}
\nocite{randich13a}
\nocite{jeffries14a}
\nocite{prisinzano16a}

The intrinsic distance modulus and reddening to the Gamma Vel cluster
have been determined as $7.76\pm 0.07$ mag and $E(B-V) = 0.038 \pm
0.012$ mag, from main-sequence fitting to the higher mass population
(Jeffries et al. 2009).  This distance is consistent with
the wider Vela OB2 association, of which the Gamma Vel cluster
appears to be a part, and also with interferometric distances to the
$\gamma^2$ Vel binary, which however appears to be younger than the 
associated PMS stars. The metallicity of the cluster was found to be
[Fe/H]=$-0.057 \pm 0.018$ from GES high resolution spectra of its
solar-type stars (Spina et al. 2014).

The GES spectra were used by Jeffries et al. (2014) and Prisinzano et
al. (2016) to confirm membership for $>200$ low-mass PMS stars
with $12<V<19$, corresponding approximately to $1.5>M/M_{\odot}>0.2$
for an assumed age of 10\,Myr.  The
catalogue and photometry of confirmed and probable members published
as table~5 in Prisinzano et al. (2016) is used here. The sample was restricted to
those stars with radial velocities between 14 and 22 km\,s$^{-1}$ and a
reliable measurement of (or upper limit to) the equivalent width of the Li~{\sc
  i}~6708\AA\ line (EWLi), in order to reduce
the possibility of contamination to just a few out of the remaining 198 stars. 
The absolute $V$ magnitude vs $(V-I)_0$ (Cousins photometry) CMD for these stars is shown in
Fig.~1a, corrected for the distance
modulus, an extinction $A_V=0.13$ and reddening $E(V-I)=0.055$ (assumed
from scaling $E(B-V)$). 
Superimposed on the data are solar-composition isochrones
taken from the Baraffe et al. (2015, hereafter BHAC15) and the Dotter
et al. (2008, hereafter Dartmouth) models,  The
Dartmouth models were transformed into colour-magnitude space with the
relationships between bolometric corrections and $T_{\rm eff}$
implied by the BHAC15 models.

\nocite{spina14a}
\nocite{baraffe15a}
\nocite{dotter08a}

The new BHAC15 models, based on a revision of the BT-Settl atmospheres
and molecular linelists described by Allard (2014), provides a much better description of the optical CMDs
of low-mass stars than previous evolutionary models. Fig.~1a shows isochrones at 7.5, 14 and 30\,Myr; there
is almost no difference between the BHAC15 and Dartmouth isochrones and
both provide an excellent match across the entire data range at an age
of $\simeq 7.5$\,Myr, estimated by minimising the median offset in $V$ between data and
isochrones for
the lowest 2/3 of the data points.\footnote{About one third
  (e.g. Fischer \& Marcy 1992) of the points in the CMD will be
   unresolved multiple systems with high enough mass ratios to cause
   them to be
  brighter and perhaps redder than their single siblings. Filtering
  these out minimises their influence on the inferred age.}
The small uncertainties in distance, reddening and placement of the
isochrone contribute to an age precision of just $\sim \pm 1$
Myr.

\nocite{fischer92a}
\nocite{allard14a}

Figure~1b shows the same set of stars in a plot of equivalent width of
EWLi vs the $V-I$ colour. The pattern of
Li abundance and therefore EWLi as a function of temperature or
equivalently, colour, is a sensitive function of mass and age in young low-mass
stars. In contracting, fully convective PMS stars, photospheric depletion commences
when the centre reaches Li-burning temperatures ($\geq 3\times
10^{6}$\,K) and convection rapidly mixes Li-depleted material to
the surface (e.g. Bildsten et al. 1997). The virial theorem tells us
that Li-burning starts when the mass-to-radius ratio increases to a
critical value (Jackson \& Jeffries 2014a); models with solar
metallicity suggest that in a
coeval group of stars, Li-burning should commence first in $\simeq
0.5M_{\odot}$ stars at $\simeq 5$\,Myr, extend to a narrow range of higher and
lower mass stars over the next 20 Myr, followed by complete depletion
for stars $<0.6 M_{\odot}$ by 120\,Myr (e.g. see Fig.~2b).

In Gamma Vel the observational data show
a rise from a small EWLi in the hotter stars towards a plateau at $V-I
\simeq 2$, but with a clear Li 
depletion ``dip'' centered at $V-I = 2.7 \pm 0.2$. There $\sim 20$ stars
at the bottom of this dip that have EWLi$<100$ m\AA, and
which must have depleted their lithium by more than 3 orders of
magnitude according to curves of growth for EWLi (see below). Superimposed on the
data we show Li depletion isochrones predicted by the same models
shown in Fig.~1a. The Li depletion factors predicted by these models have
been converted from Li abundances into EWLi assuming an initial Li abundance of
$A({\rm Li}_0)=3.3$ and by folding the predicted $A({\rm Li})$, $T_{\rm eff}$ coordinates
at each mass and age 
through LTE curves of growth for the Li\,{\sc i}~6708\AA\ feature,
defined by Soderblom et al. (1993) for
$T_{\rm eff}>4200$\,K and by Zapatero-Osorio et al. (2002) for $T_{\rm
  eff}<4200$\,K.\footnote{The predicted
  EWLi for a given Li abundance and $T_{\rm eff}$ from the
  Zapatero-Osorio et al. estimates are within 10 per cent of those
  independently calculated by Palla et al. (2005) at $T_{\rm eff}\leq 3600$\,K.} The colours are from the models as for Fig.~1b and
reddening has been applied.

\nocite{palla05a}
\nocite{soderblom93a}
\nocite{zapatero02a}

Both sets of models (BHAC15 and Dartmouth) are quite 
similar and {\it cannot fit the Li depletion data at all}. At
the age of 7.5\,Myr implied by the CMD there is far too little Li
depletion (by orders of magnitude in Li abundance and factors of
several in EWLi). An age of $>14$ Myr is needed to provide the required
Li depletion, but this would still occur at colours that are far too
blue by $>0.5$ mag. 
Only a small part of this age discrepancy could be
explained by unresolved binarity making some of the Gamma Vel stars
redder -- the most likely candidate binaries, those well above the
7.5 Myr isochrone in Fig.~1a, are indicated. In
Appendix~A (see Fig.~A1 a,b) it is shown that exactly the same result is arrived at
from the $V$ vs $V-K$ CMD and EWLi vs $V-K$ diagram. Neither can the
additional Li depletion be explained by an anomalous metallicity.
Although Li depletion {\it is} metallicity sensitive, in the sense that
greater depletion is expected in more metal-rich stars of a given mass,
Spina et al. (2014) have established that the metallicity of the Gamma
Vel cluster is slightly subsolar (see the Introduction).

In summary, the most commonly used, contemporary low-mass
stellar evolution models are unable to simultaneously describe the CMD
and Li depletion pattern of PMS stars in the Gamma Vel cluster.

\section{Models with Inflated Radii} \label{inflation}

If stars of a given mass and age have larger radii and cooler $T_{\rm
  eff}$ than predicted by
the standard models, then their central temperatures will be lower and
their Li depletion will be delayed; however that depletion will
occur in stars with redder colours. This, along with the context
described in the Introduction, leads to consideration of whether radius inflation 
could solve the problem identified in Section~2.

\begin{figure*}
\centering
\includegraphics[width=160mm]{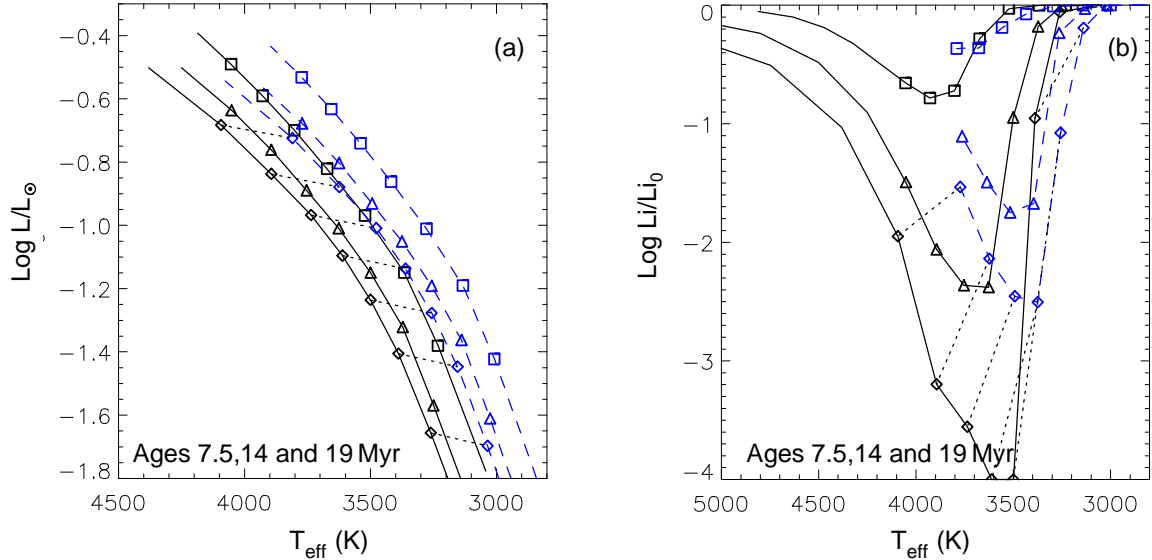}
\caption{The effects of a 10 per cent radius inflation ($\beta=0.25$). 
  The left hand panel shows isochrones in the
  Hertzsprung-Russell diagram
  from the BHAC15 model at 7.5 (diamonds), 14 (triangles) and 19 Myr
  (squares).
  The blue dashed lines show
  the same isochrones, modified for the effects of radius
  inflation. Mass points
  from $0.2\,M_{\odot}$ to $0.8\,M_{\odot}$ in $0.1\,M_{\odot}$ steps
  are indicated by the open symbolds on each isochrone. A set of black dotted lines indicate (for
  the 19 Myr isochrone) how the effects of inflation modify each mass
  point. The
  inflated 19 Myr isochrone lies very close to the 7.5\,Myr BHAC15 isochrone.
  The right hand panel shows the same isochrones in the Li
  abundance versus $T_{\rm eff}$ plane. These diagrams illustrate that for a
  given mass and age, the
  effect of radius inflation is to reduce the luminosity, lower the surface
  temperature and reduce the amount of Li depletion.}
\end{figure*}

The BHAC15 models predict that stars with $M<0.7\,M_{\odot}$, 
deplete their lithium whilst descending near-vertical, fully-convective
Hayashi tracks in the HRD; the
luminosity of the star results principally from the release of
gravitational energy. In this case a simple polytropic model may be
used to investigate the effects of reduced heat transfer in the surface
layers of a star on the evolution of stellar radius, luminosity and
Li abundance. 
As discussed in the Introduction, a number of authors have experimented
with evolutionary models that incorporate the effects of rapid rotation
and magnetic
fields, either in the form of inhibited convection throughout the star,
or the blocking of flux at the surface by cool starspots. It is well
known that low-mass PMS stars (including those in Gamma Vel) do
rotate rapidly and show high levels of magnetic activity (Frasca et
al. 2015). Here, we 
remain agnostic about which, if any, of these mechanisms might be
effective in low-mass PMS stars, but hypothesise that {\it something}
causes them to be larger at a given mass and age by reducing the
radiative flux at the surface.

\nocite{frasca15a}

Jackson and Jeffries (2014a) considered the effect of 
starspots reducing the luminosity of PMS stars by
a factor $(1-\beta$) relative to the luminosity of an unspotted star at
the temperature of the unspotted photosphere. Equations 2 and 3 of that
paper show that the effect of starspots is to reduce the luminosity,
$L$ whilst increasing the radius, $R$ of a star relative to an unspotted
star of similar mass $M$ and age $t$ such that:
\begin{equation}
\left[\frac{R}{R_0}\right]_M \propto (1 -\beta)^{-D} t^{-D}\, ,
\end{equation}
\begin{equation}
\left[\frac{L}{L_0}\right]_M \propto (1- \beta)^D t^{D-1}\, ,
\end{equation}
where $D = (A-4)/(3A-4)$ and $A = \partial \log L/\partial \log T_{\rm
  eff}$ at fixed mass. For near-vertical Hayashi
tracks $A$ is very large and the exponent $D \simeq 1/3$, i.e.  the radius at a
given age increases with spot coverage as $(1-\beta)^{-1/3}$, while the
  luminosity decreases in equal proportion. It follows that, for a
  fixed mass, the age at which the central temperature ($\propto M/R$) 
  reaches the ignition point  of lithium (see
  Bildsten et al. 1997) is increased by a factor $(1-\beta)^{-1}$.

\nocite{bildsten97a} 

Whilst this model of reduced heat transfer at the stellar surface was
used to investigate the effect of starspots it can equally be used to
model other effects that inflate stars on Hayashi tracks at a
given $M$ and $t$. Specifically, it can approximately represent the
effects of reducing the efficiency of convective heat transfer due to
interior magnetic fields or some other reduction in the
convective mixing length provided that the majority of the star remains
adiabatic. This is expected to be true for young PMS
stars, since super-adiabatic layers are limited to quite close to the
stellar surface (see fig.~10 in Feiden \& Chaboyer 2014). In this case $\beta$ is simply
related to the relative increase in radius, $R/R_0$ at fixed
$M$ and $t$ as $\beta = 1- (R/R_0)^{-3}$, i.e. a 10 per cent
increase in radius corresponds to $\beta=0.25$.

In Fig.~2a we show how this polytropic model modifies the 
BHAC15 isochrones and the positions of stars of a given mass
in the HRD for $\beta=0.25$.
The inflated models are calculated for masses
$0.2<M/M_{\odot}<0.8$ where the stars are (almost) fully convective at
the ages considered here. 
Stars of a given mass and age
become slightly less luminous and significantly cooler. As a
result, any age inferred from matching these models to data in the CMD
would be significantly older and a mass estimated from position in the HRD would be larger
than estimated from standard models.

In Fig.~2b we show how the same polytropic model affects isochrones in
the Li abundance-$T_{\rm eff}$ plane. These are calculated by assuming
that the same relationship between $M/R$ (and hence central temperature)
and Li depletion holds for both standard and inflated models.
Inflation increases the radius at a given mass and age and hence
decreases the central temperature. A star of a given mass takes longer
to reach the Li-burning stage, but has a cooler $T_{\rm eff}$ when
it does so. The result is {\it less} Li depletion at a given age, but
the centre of the ``Li-dip'' becomes cooler by $\sim 300$\,K.

\begin{figure*}
\centering
\includegraphics[width=160mm]{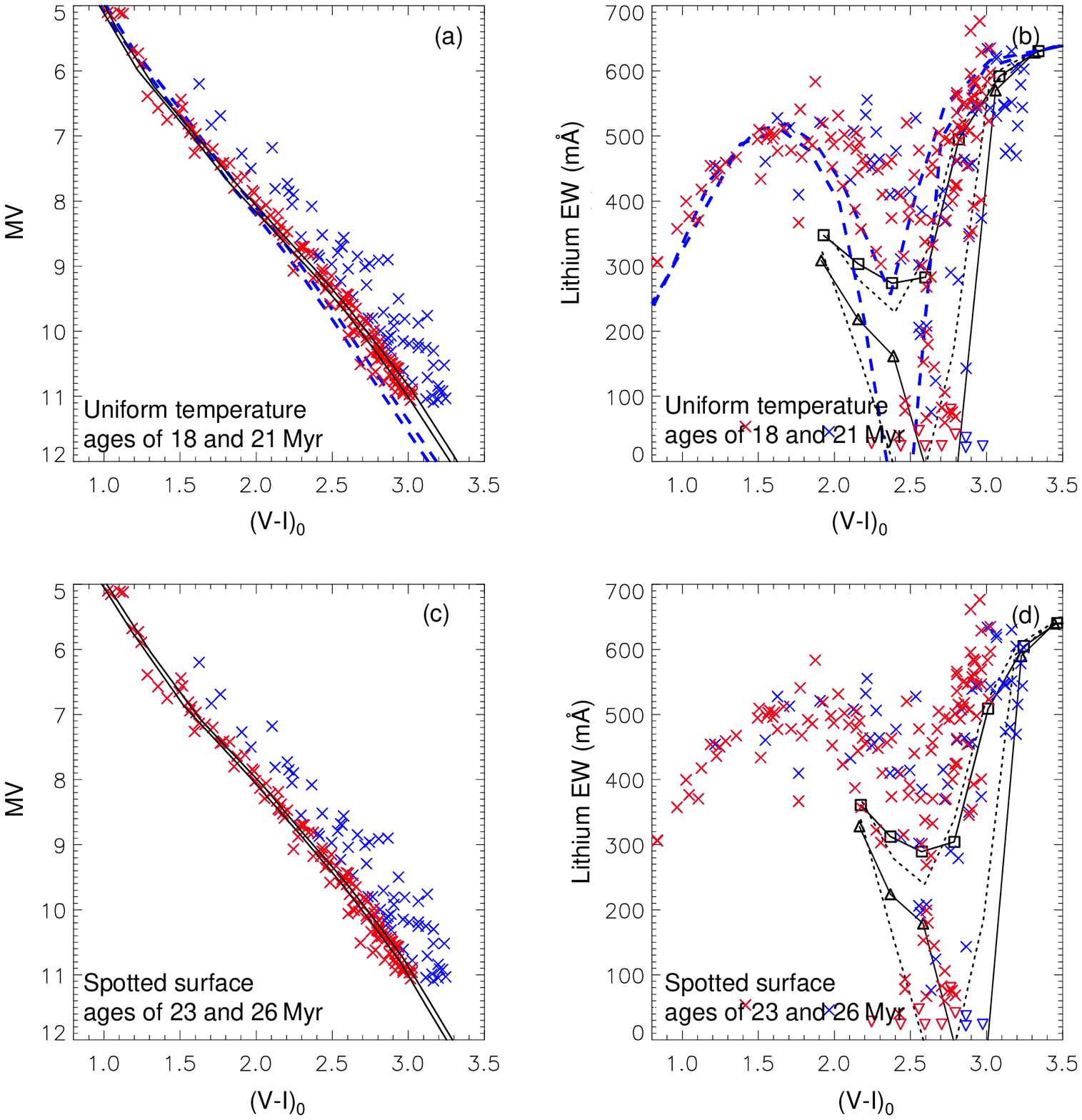}
\caption{
The absolute magnitude versus intrinsic colour diagram and lithium depletion pattern
  in the Gamma Vel cluster (same data and symbol colour meaning as for Fig.~1) 
  compared with isochrones at 18 and 21\,Myr modified by
  radius inflation, with mass points from $0.2\,M_{\odot}$ to
  $0.8\,M_{\odot}$ indicated by open symbols. Panels (a) and (b) refer to the uniform temperature
  radius inflation with $\beta=0.25$, while panels (c) and (d) are
  similar diagrams for a spotted model with $\beta=0.41$ and a spotted
  to unspotted temperature ratio of 0.85.
  Black solid lines and black short-dashed lines (almost coincident
  with the black solid lines in panels (a) and (c) and hence not visible) are  
  BHAC15 and Dartmouth isochrones modified by
  inflation according to the
  polytropic model described in Section~3 at the ages indicated in the plots.
  In panels (a) and (b) the blue dashed lines shows 18 and 21 Myr
  isochrones from the magnetic Dartmouth models.
}
\end{figure*}

\section{Comparison of models with inflated radii to the Gamma Vel cluster}

Whilst the relationship between $\beta$, radius inflation and Li
depletion is to first order independent of whether the inflation is
caused by starspots or something else, the consequent {\it appearance}
of the star is not. The {\it effective} temperature is reduced by a
factor $(1-\beta)^{1/4}$. For a star with uniform surface temperature, the
effect on $V$ and $V-I$ is calculated straightforwardly using the
relationships between luminosity, temperature, absolute magnitudes and
colours implied by the BHAC15 models (ignoring the small gravity
change). For spotted stars the light comes from both spotted and
unspotted regions. The magnitudes and colours will depend on $\beta$
{\it and} the ratio of the temperatures of the spotted and immaculate
photosphere. The procedure for calculating this is described in Appendix~B.

In Figs. 3a,b inflated (by 10 per cent at a given
mass and age) versions of isochrones from the BHAC15 and Dartmouth
models are compared with the Gamma Vel cluster data.
The distance modulus and reddening are as
described in Section 2. Li abundances are converted to EWLi using the
same curves of growth described in Section~2.
Figure 3a shows that for 10 per cent inflation, an 
18-21 Myr isochrone is now an excellent match to the CMD using either the
inflated BHAC15
or inflated Dartmouth models. This is significantly older than found in
Section 2 because stars of a given mass and age are cooler in the inflated models.
Figure 3b shows 18 and 21\,Myr Li depletion isochrones for these inflated models,
illustrating how extremely age-dependent the
Li depletion is at this epoch. A $\sim 20$ Myr isochrone is now a much better fit,
both to the depth and location of the Li-dip in the Gamma Vel data, than was
seen in Fig.~1b. The
inflated BHAC15 and Dartmouth models are similar in terms of
predicted Li-dip depth, but the inflated Dartmouth models fare 
slightly worse in that the colour of their predicted Li-dip is still
slightly blueward of the observations. There is
no {\it a priori} reason to suppose that one $\beta$ is
appropriate for stars at all masses, but a uniform value does seem
to work reasonably well.

Uncertainties on the required values of $\beta$ and consequent ages
were estimated in the following way. The
required age was determined from the EWLi-colour plot. An
initial $\beta$ value was chosen, which defined
a narrow range of isochrones that provided a minimum in EWLi somewhere between
300m\AA\ and undetectable. The centre of this age range in turn defines a
new $\beta$ value (and hence level of radius inflation) 
that will provide a match to the CMD using the same procedure
used to define the CMD age in Section~2. This procedure is then
iterated using the new $\beta$ value and converges on the
parameters and age ranges shown in Table~1. The colour of the Li-dip is not used as a
fitting criterion. The finite precision of the age estimate inferred from the
non-magnetic models ($7.5 \pm 1$ Myr) formally gives an uncertainty of
$\pm 0.04$ in the final $\beta$ estimates, but leads to only an
additional 1 Myr uncertainty in the age (which is ignored).

\begin{table}
\caption{Best fitting ages for inflated models}
\begin{tabular}{lccccc}
\hline
Type of & $\beta^{a}$ & Radius    & Temperature     & Spot     & Age \\
model   &         & inflation$^b$ &     ratio$^c$    & coverage$^d$ & (Myr)\\
\hline
Uniform &  0.25   & 10\%      & 0.93             & --        & 18--21 \\
Spots   &  0.30   & 13\%      & 0.90             & 0.87      & 19--22 \\
Spots   &  0.41   & 19\%      & 0.85             & 0.86      & 23--26 \\
Spots   &  0.51   & 27\%      & 0.80             & 0.86      & 28--32 \\
\hline
\multicolumn{6}{l}{$a$ The fraction of flux blocked at the surface.} \\
\multicolumn{6}{l}{$b$ The level of inflation compared to non-magnetic models.}\\ 
\multicolumn{6}{l}{$c$ For uniform models this is the ratio of $T_{\rm
                       eff}$ to the $T_{\rm eff}$ of an uninflated} \\
\multicolumn{6}{l}{star of the same mass. For the
                       spotted models this is the ratio of spotted} \\
\multicolumn{6}{l}{to unspotted temperatures
                       $T_{sp}/T_s$ -- see Appendix B.} \\
\multicolumn{6}{l}{$d$ The surface coverage fraction of cooler spots.} 
\end{tabular}
\end{table}

Figures~3a,b also show
equivalent isochrones at 18 and 21 Myr predicted
by the ``magnetic Dartmouth models'' described in Feiden \& Chaboyer
(2014) and Feiden (2016). These implement magnetic inhibition of
convection constrained by a boundary condition of an average 2.5 kG
equipartition field at the stellar surface. At a given mass, and an age
of $\sim 20$\,Myr,
these models predict that stars are inflated by between 12 per
cent at $0.8\, M_{\odot}$ and 7 per cent at $0.2\,M_{\odot}$. These
models are quite close to our simple polytropic models in the
CMD, but are too blue at the lowest masses.  
In the EWLi versus colour diagram, the depth and colour of the
Li-dip are similar to our simple models in Fig.~3b, 
perhaps providing a worse
fit to the Gamma Vel data on the red (low-mass) side of the Li-dip, 
but a better fit on the blue side.
The former discrepancy is likely due to the lower level of
inflation at lower masses than assumed in our best fitting uniform
temperature and $\beta$ polytropic models. The latter is because
at $\sim 20$\,Myr, the magnetic Dartmouth models have a small radiative
core at $M \geq 0.6M_{\odot}$, meaning that the base of the convection
zone 
is cooler than the core; this reduces
the amount of photospheric Li depletion over our simple fully
convective polytropic approximation, better matching the data.

The models shown in Figs.~3a,b assume that the inflated stars have a
uniform surface temperature, but the polytropic model can equally
describe inflation due to dark starspots (see Section 2).
We find that the reddening of the isochrones is maximised
for the case of a uniform temperature; spotted models produce a
smaller redward shift for the same $\beta$ and 
level of inflation. This means that
if the radius inflation is caused by spots then agreement between the CMD age
and lithium depletion ages requires a larger $\beta$, more radius
inflation and older ages than for a uniform
temperature reduction. Consequently, spotted models require a very large
spot coverage if they are to be the sole cause of the radius inflation.
Results for uniform temperature reductions and
spotted models are summarised in Table~1, and Figs.~3c,d show examples of
inflated models that match the CMD and Li depletion pattern
for an assumed spot temperature ratio of $T_{sp}/T_{s}=0.85$ (see Appendix~B).

\section{Discussion}

In Section~2 we showed that the most commonly used evolutionary models
cannot simultaneously describe the CMD and Li depletion pattern of the
Gamma Vel cluster at a common age. Ages based on Li appear to be much
older, though we note that these models do not describe Li depletion at
{\it any} age, because they predict Li depletion occuring at much bluer
colours than observed.  Based on what is known about the radii of
magnetically active stars in eclipsing low-mass binaries, it is natural
to seek a resolution in terms of modifying the mass-radius relationship
at low-masses. Figures~3 and A1 show that a resolution is possible if
the radii of stars are inflated with respect to standard models by
about 10 per cent at a given mass and age.  The results of a simple
polytropic model are in remarkably good agreement with the ``magnetic''
Dartmouth models of Feiden \& Chaboyer (2014) for $M\leq 0.6M_{\odot}$,
which show similar levels of inflation and are also a reasonable match
to the data.  Starspots could also be responsible for radius
inflation. Our modelling (see Table~1) suggests that if they were to be
the {\it sole} cause of the discrepancy between HRD/CMD ages and
Li-depletion ages, then this would require significantly more radius
inflation and an unrealistically high level of spot coverage -- spot
modulation and doppler imaging can give only lower limits to spot
coverage, but modelling of TiO bands in the spectra of magnetically
active stars suggets 50 per cent coverage may be possible (e.g. O'Neal
et al. 2004 and extensive discussion in Jackson \& Jeffries 2014b).
In reality, these active stellar photospheres are more likely to feature
a distribution of temperatures rather than the simple uniform or bi-modal
extremes assumed by our models and the radius inflation would be due
to a combination of magnetically moderated effects.

\nocite{oneal04a}

Yee \& Jensen (2010, see also Malo et al. 2014) 
noted a quantitatively similar discrepancy between
ages from the HRD and ages inferred from Li depletion for stars
in the young Beta Pic moving group. They also suggested that radius
inflation might be a way to resolve the difference, noting that
inflation would increase ages from the HRD and would shift Li
depletion to cooler $T_{\rm eff}$. However, they suggested
that inferred ages from Li depletion would be {\it reduced} by radius
inflation. In fact as we have shown here, and has also been
demonstrated by Jackson \& Jeffries (2014a) and Somers \& Pinsonneault
(2014, 2015a), radius inflation {\it delays} the onset of Li depletion
leading to {\it older} inferred ages. As a result, the ages from
the CMD/HRD and Li-depletion can only be reconciled at ages at least a
factor of two
older than inferred from the CMD/HRD using standard models. 
This assumes
that the inflationary factors were already in place prior to
commencement of Li burning. If magnetic activity is the
culprit then this seems reasonable, since high levels of magnetic
activity are ubiquitously seen in much younger PMS stars than those
considered here. 

Almost identical conclusions were reached for the Beta Pic group by
Messina et al. (2016). They showed that a version of the magnetic
Dartmouth models was able to describe the CMD and Li-depletion
patterns at a common age of $25\pm 3$ Myr. This is entirely consistent
with our analysis and age of 18-21 Myr for the Gamma Vel cluster using
similar models -- an empirical comparison of the Li
depletion patterns in the two datasets (compare Fig.~A1 with fig.~1
from Messina et al.) shows that Gamma Vel is definitely younger than Beta Pic;
the lithium-dip is much wider in Beta Pic (by about 0.5 mag in $V-K$), 
and extends to significantly bluer
colours. However, given the extreme age sensitivity of Li depletion, an
age difference of a few Myr is all that is required.

\nocite{malo14a}

Radius inflation due to magnetic activity has the capacity
to explain a number of observational puzzles that have emerged in
recent years concerning young stars. The increase in age that would
follow from using models with radius inflation solves a long-standing
discrepancy between the ages derived from the HRD/CMD of low-mass PMS
stars using standard models and the considerably older ages determined
using the ZAMS turn-on or nuclear turn-off ages from higher mass stars
with little magnetic activity (e.g. Naylor 2009; Pecaut et
al. 2012). Bell et al. (2013) found that by adopting empirical
relationships between bolometric correction and colour derived using a
fiducial sequence of (magnetically active) 
ZAMS stars in the Pleiades cluster, they 
could bring the ages derived
from high- and low-mass stars in young clusters into agreement. The
empirical modification to the bolometric corrections is equivalent
to changing the radius of the low-mass stars at a given $T_{\rm
  eff}$. More recently, Feiden (2016) showed that the magnetic
Dartmouth models (discussed in Section 3) inflated the radii and 
increased the derived ages of
low-mass PMS stars in the Upper Sco association by a factor of two, 
bringing them into
agreement with ages determined from higher mass stars.

\nocite{naylor09a}
\nocite{pecaut12a}

The large age increases suggested here apply to young clusters for which the age
has been previously determined using isochronal fits to their 
low-mass PMS stars. 
Ages derived from the ``lithium depletion boundary'' (LDB) -- the {\it
  luminosity} below which Li remains unburned in the low-mass stars of
clusters with ages from 20--130\,Myr (e.g. Bildsten et al. 1997), will
be considerably less affected. LDB ages have been proposed as the most
model-independent of age determination methods in young clusters
(Soderblom et al. 2014) and indeed Jackson \& Jeffries (2014a) showed that
these ages are only increased by $\sim (1-\beta)^{-0.5}$ due to magnetically
inflated radii. Thus a 10 per cent radius inflation requiring
$\beta=0.25$ would increase the determined LDB age by just 15 per
cent. For example Binks \& Jeffries (2016) showed that adopting the
inflated Dartmouth magnetic models discussed here increased
the LDB age of the Beta Pic moving group from 21\,Myr to 24\,Myr.

\nocite{binks16a}

There are also
discrepancies between model predictions and the measured radii and masses
for low-mass PMS and ZAMS eclipsing binary systems in clusters and
associations (Kraus et al. 2015, 2016; 
David et al. 2016; Sandberg Lacy et al. 2016).  On the basis of position in the HRD, 
evolutionary models predict masses that are too low and ages that are
much younger than suggested by higher mass stars in the same
clusters/systems.  
Similarly, the dynamical masses of astrometric PMS binaries are higher
than implied by their measured luminosities and standard evolutionary models
(Hillenbrand \& White 2004; Azulay et al. 2015; Rizzuto et al. 2016). Figure~2a
demonstrates that these phenomena would be expected if stars were
inflated by $\sim 10$ per cent with respect to the standard evolutionary models -- $T_{\rm
  eff}$ is reduced by 7 per cent, inferred
ages are increased by a factor of $\geq 2$, whilst masses inferred from
the HRD/CMD are
almost doubled at the lowest masses considered here (from $0.2M_{\odot}$
to $\sim 0.35M_{\odot}$), with a smaller effect at higher masses (from
$0.6M_{\odot}$ to $\sim 0.8M_{\odot}$). If masses were estimated from
luminosities (or absolute magnitudes), the effect would be much smaller.

\nocite{sandberg16a}

Finally, an explanation of 
the well known issue of a rotation-dependent dispersion of Li
abundance at a given $T_{\rm eff}$ in ZAMS stars (e.g. Soderblom et
al. 1993; Barrado et al. 2016) has been proposed by
Somers \& Pinsonneault (2014, 2015a,b), which 
assumes that stars have radii that are inflated by differing amounts
depending on their level of rotation and magnetic activity. This leads
to differing levels of Li depletion, as explained in Section 3, and a
dispersion in Li abundances at the ZAMS that was imprinted during PMS
evolution. This dispersion is important because it illustrates the need
for a solution that can vary from star-to-star, rather than a global
change to evolutionary models that would affect all stars of the same
mass equally. For instance, similar radius inflation could be produced
by artificially reducing the convective efficiency or altering the adopted
relationship between temperature and optical depth in the atmosphere
(Chen et al. 2014), but this could not explain any dispersion in stellar properties. 

\nocite{chen14a}
\nocite{barrado16a}
\nocite{bouvier16a}

A considerable dispersion in EWLi at $V-I \sim 2.7\pm 0.2$ {\it is}
present in the Gamma Vel data and cannot be explained by uncertainties
in EWLi alone\footnote{The median uncertainty in EWLi is
15\,m\AA, see fig.~5 in Prisinzano et
al. (2016)}. The presence of stars with undetectable levels of Li
  can also not be explained in terms of contamination by non-members;
  there are too many with a radial velocity close to the cluster
  centroid and all have some secondary indication of a youthful nature
  -- see Prisinzano et al. (2016) for details. If this dispersion were present {\it at a given $T_{\rm
    eff}$ or colour}, it would favour an explanation in terms of
differences in $\beta$ caused by differing 
magnetic activity levels or rotation rates from star-to-star.
Here, the simplified assumption has been made of a single value of
$\beta$ across the mass range, which could perhaps be
justified in terms of the consistent (saturated) levels of
chromospheric activity exhibited by PMS stars in Gamma Vel despite a
spread in rotation rates (see Frasca
et al. 2015).  On the other hand, it is uncertain how closely the levels of
spot coverage or interior magnetic field strengths are correlated with
chromospheric activity -- a Li-depletion-rotation connection has recently been
identified in the similar PMS stars of NGC 2264 (age 5--10\,Myr) by Bouvier et al. (2016).
Attempts to investigate this in Gamma Vel are hampered by a lack of rotation period
data and by both intrinsic and measurement uncertainties. Photometric
precision, variability, small reddening differences, differences in the
spot coverage when the photometry was performed, the presence of
unresolved binary companions and the possibility of a small age spread
in the population (see Jeffries et al. 2014) can all contribute to a
scatter in $V-I$ and EWLi that could obscure any relationship between
Li depletion and rotation/activity over such a narrow Li-dip. The
nature of the EWLi dispersion in Gamma Vel and in other young clusters
observed by GES is deferred to another paper.

A remaining puzzle is the status of the massive binary $\gamma^2$~Vel,
which appears to be at the centre of the cluster. According to
calculations by Eldridge (2009), the binary has an age of $5.5\pm
1$\,Myr. This age discrepancy between the $\gamma^2$~Vel and the PMS
star surrounding it was already noted by Jeffries et
al. (2009, 2014). The introduction of inflated
models for the PMS stars makes the discrepancy larger. Data from
the Gaia satellite should ultimately settle whether $\gamma^2$ Vel and the
lower mass stars are physically related or are merely a line of sight
coincidence.

\nocite{kraus16a}
\nocite{eldridge09a}

\section{Summary}

The colour magnitude diagram of low-mass PMS stars in the Gamma Vel
cluster is well-fitted with
isochrones from the most commonly used evolutionary models at 
an age of $7.5 \pm 1$ Myr. However, the same models cannot explain the pattern of
Li depletion in the same stars at the same age - 
there is too little Li depletion and it occurs at colours that are too blue.
A simple polytropic model that simulates the effects of starspots or magnetic
fields in inhibiting the radiative flux at the surface of a fully
convective star is used to modify the evolutionary models, showing that
stars become larger, less luminous and cooler at the same mass and age. These
revised models are able to simultaneously fit the CMD and the Li
depletion pattern -- at a significantly increased age of 18--21\,Myr if radii
are inflated by 10 per cent and the
surface temperatures are uniform, or at even older ages if the radii
are solely inflated due to a very large covering fraction of cool starspots.
These conclusions are very similar to those arrived at by Messina et
al. (2016) who performed a combined analysis of the CMD and Li
depletion pattern for stars in the slightly older Beta Pic moving group.

The inflation of radii by magnetic activity has the capacity to resolve
a number of astrophysical problems for low mass PMS stars, including:
discrepancies between measured masses, radii, luminosities and
temperatures of PMS binary systems; discrepancies between the ages
determined from the HRD/CMD for low-mass PMS stars and their higher
mass siblings; and the patterns and dispersion of Li depletion seen
among low-mass PMS and ZAMS stars. This scenario would require an
increase in the timescales of PMS evolution of at least a factor of
two, and since observational constraints on timescales for the
evolution of circumstellar disks and planetary formation are keyed in
to these PMS timescales, they would have to be revised upwards by a similar
factor. In addition, the estimation of stellar masses from 
positions in the CMD/HRD would be significantly affected. PMS stars that are
inflated by magnetic activity would have masses that are
larger than implied by standard evolutionary models. 

\section*{Acknowledgements}

RDJ and RJJ acknowledge support from the UK Science and Technology
Facilities Council (STFC).  Based on data products from observations
made with ESO Telescopes at the La Silla Paranal Observatory under
programme ID 188.B-3002. These data products have been processed by the
Cambridge Astronomy Survey Unit (CASU) at the Institute of Astronomy,
University of Cambridge, and by the FLAMES/UVES reduction team at
INAF/Osservatorio Astrofisico di Arcetri. These data have been obtained
from the Gaia-ESO Survey Data Archive, prepared and hosted by the Wide
Field Astronomy Unit, Institute for Astronomy, University of Edinburgh,
which is funded by the STFC. This publication makes use of data
products from the Two Micron All Sky Survey, which is a joint project
of the University of Massachusetts and the Infrared Processing and
Analysis Center/California Institute of Technology, funded by the
National Aeronautics and Space Administration and the National Science
Foundation.

This work was partly supported by the European Union FP7 programme
through ERC grant number 320360 and by the Leverhulme Trust through
grant RPG-2012-541. We acknowledge the support from INAF and Ministero
dell' Istruzione, dell' Universit\`a' e della Ricerca (MIUR) in the
form of the grant ``Premiale VLT 2012''. This research was partially
supported by INAF through a PRIN-2014 grant. The results presented here
benefit from discussions held during the Gaia-ESO workshops and
conferences supported by the ESF (European Science Foundation) through
the GREAT Research Network Programme.

\bibliographystyle{mn2e} 
\bibliography{refs}

\appendix

\section{The $V-K$ colour}

\begin{figure*}
\centering
\includegraphics[width=160mm]{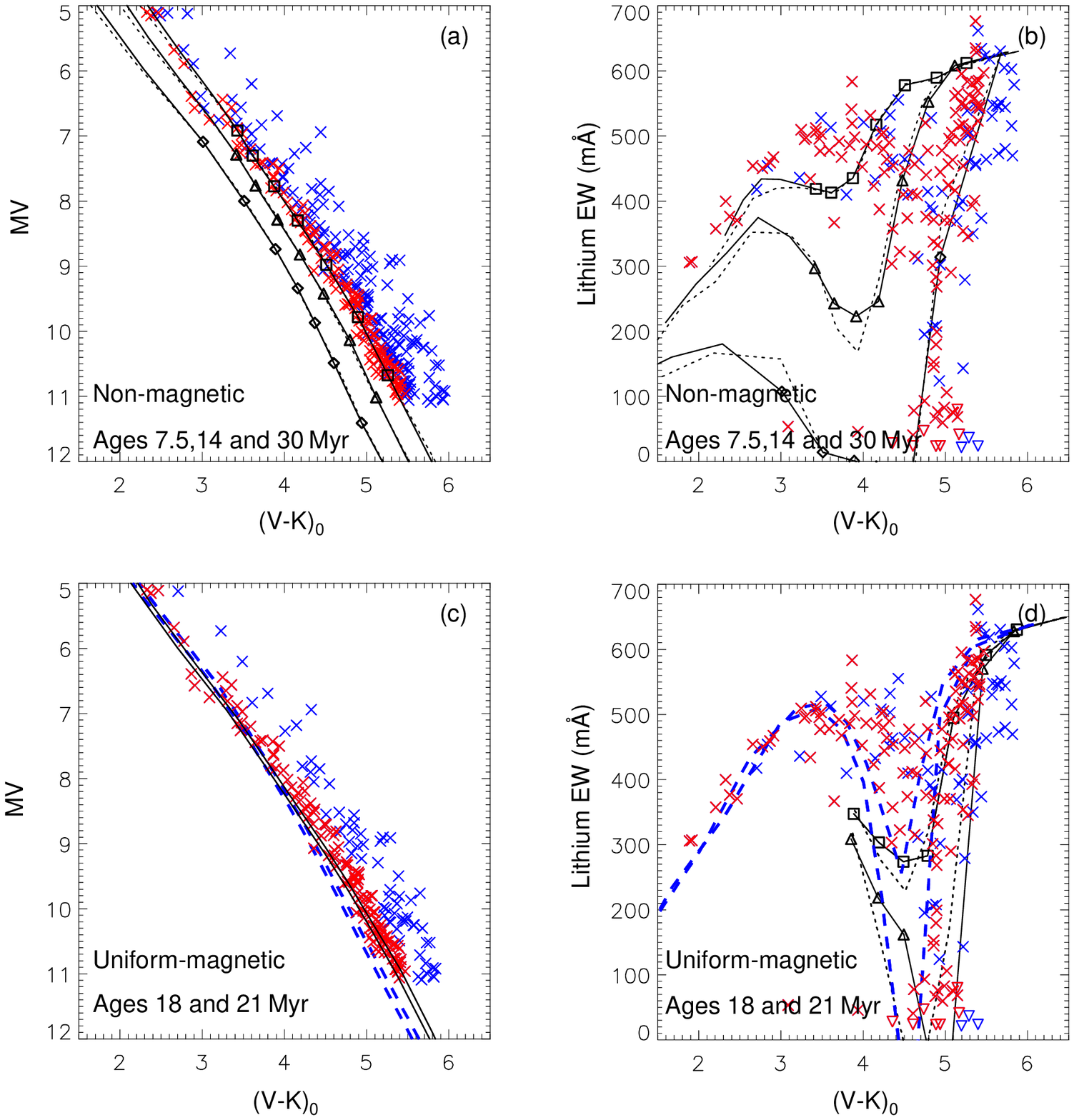}
\caption{
The absolute magnitude versus intrinsic $V-K$ diagram and lithium depletion pattern
  in the Gamma Vel cluster (same data and symbol colour meaning as for Fig.~1) 
  compared with isochrones calculated using the $V-K$ (2MASS) colour and
  reddening $E(V-K)=0.113$. 
  Plots (a) and (b) are similar to Fig.~1.
  Black solid lines and black short-dashed lines 
  are BHAC15 and Dartmouth isochrones respectively at 7.5, 14 and
  30\,Myr. Plots (c) and (d) show 18 and 21\,Myr BHAC15 and Dartmouth isochrones 
  (solid and dotted) modified for 10 per cent radius
  inflation according to the
  polytropic model described in Section~3, assuming a uniform
  surface temperature. The blue dashed lines show the same isochrones
  for the magnetic Dartmouth model.
}
\end{figure*}

The main plots and conclusions of the paper have been checked using the
$V-K$ colour in addition to $V-I$. The equivalents to Figs.~1 and 3a,c
are shown in Fig.~A1. The $K$ magnitudes here are on the 2MASS system
(from Skrutskie et al. 2006) and the model $V-K$ colours on the CIT
system were transformed to the 2MASS system by subtracting 0.024 mag
(Carpenter 2001).
The isochrones shown are not fitted but placed
at the same ages that are discussed in the main text for the $V-I$
isochrones. The reader can readily see that the problem outlined in
Section~2 is also apparent when using $V-K$; Figs.~A1a,b
show that an age of
7.5\,Myr fits the CMD (though a slightly younger age might be better),
but the same (or any older) isochrone in the EWLi vs $V-K$ does not
match the data. Applying the same 10 per cent level of inflation  to
the BHAC15 and Dartmouth models at a given mass yields the solid and
dotted isochrones
plotted in  Figs.~A1c,d which provide
an acceptable fit to both diagrams at an age of $\sim 20$\,Myr. The
magnetic Dartmouth models yield an inferior fit at this age, being too
blue at lower masses, but qualitatively address the problem in the same way.

\nocite{carpenter01a}
\nocite{skrutskie06a}

\section{The effect of starspots on magnitude, colour and lithium
  equivalent width}
A simple two-temperature model is used to determine the effect of
starspots on the observable properties of spotted stars (see Jackson \&
Jeffries 2014b). 
Starspots affect the magnitude and colour of stars on the Hayashi track in two ways; 
\begin{itemize}
	\item There is a small increase in the temperature {\it of the unspotted
     photosphere} of the spotted star, $T_s$ relative to that of an unspotted star, 
     $T_u$ of the same mass and age. From Eqs.~(1) \& (2)
\begin{equation}
(T_s/T_u)_{M,t} = (1-\beta)^{\tiny{-(3D-1)/4}}\ . 
\end{equation}
     In practice $D \sim 1/3$, so $T_s$ is only slightly larger than $T_u$.
  \item There is also a contribution to the observed stellar flux from 
    the spotted area of the photosphere that always makes the star appear redder. 
    The relative magnitude of this contribution depends on the spot temperature $T_{sp}$
    and the fraction of the surface covered by starspots,
    $\gamma=\beta/(1-T_{sp}^4/T_s^4)$
\end{itemize}

\noindent{The $V$-band flux is the sum of the flux from the unspotted surface of
area $(1-\gamma)4\pi R^2$ and temperature $T_s$, and the flux from the spotted
surface of area $\gamma 4\pi R^2$ and temperature $T_{sp}$; giving a combined
$V$-band bolometric correction of;}
\begin{equation}
	{\rm BC}_{V,{\rm star}} = 2.5\log \left(\frac{1-\gamma}{1-\beta}10^{\tfrac{{\rm BC}_{V,T_s}}{2.5}}+	
	\frac{\gamma(\tfrac{T_{sp}}{T_s})^4}{1-\beta}10^{\tfrac{
            {\rm BC}_{V,T_{sp}}}{2.5}}\right)\, ,
\end{equation}
where BC$_{V,T}$ is the bolometric correction at temperature $T$ given by the BHAC15 model 
at the appropriate stellar age.  A similar expression is used to evaluate
the $I$- and $K$-band bolometric corrections in order to determine the
$V-I$ or $V-K$ colours of the spotted star.

EWLi is also calculated from the curves of growth discussed in
Section~2, assuming that both spotted and unspotted areas have the
same Li abundance,  and that the EWLi contribution, corresponding to the
assumed temperatures of the spotted and unspotted regions,
is weighted as $\gamma T_{sp}^4$ and $(1 -\gamma)T_s^4$
respectively. This in turn assumes that the continuum flux at
6708\AA\ scales roughly as the bolometric flux. The exact weighting
makes little difference since the EWLi vs $T$ relation for a given
abundance is quite flat.

\label{lastpage}


\bsp 
\end{document}